\documentclass[conference]{IEEEtran}
\usepackage[english]{babel}
\usepackage[utf8x]{inputenc}
\usepackage{amssymb}
\usepackage{amsmath}
\usepackage{graphicx}
\usepackage[]{algorithm}
\usepackage{rotating}
\usepackage{multicol}
\usepackage{algpseudocode}
\usepackage{esvect}
\usepackage{relsize}
\usepackage{booktabs}
\usepackage{multirow}
\usepackage{siunitx}

\begin{document}
\title{Seismic Imaging: An Overview and Parallel Implementation of Poststack Depth Migration}
\author{\IEEEauthorblockN{
Ahmad~Shawahna\IEEEauthorrefmark{1},
Syed~Abdul~Salam\IEEEauthorrefmark{2}, and
Mayez~Al-Mouhamed\IEEEauthorrefmark{1}
}
\IEEEauthorblockA{\IEEEauthorrefmark{1}Department of Computer Engineering\\}
\IEEEauthorblockA{\IEEEauthorrefmark{2}Department of Electrical Engineering\\}
King Fahd University of Petroleum and Minerals, Dhahran-31261, KSA\\
\{g201206920, g201302650, mayez\}@kfupm.edu.sa
}

\maketitle

\begin{abstract}
Seismic migration is the core step of seismic data processing which is important for oil exploration. Poststack depth migration in frequency-space ($f-x$) domain is one of commonly used algorithms. The wave-equation solution can be approximated as FIR filtering process to extrapolate the raw data and extract the subsurface image. Because of its computational complexity, its parallel implementation is encouraged. For calculating the next depth level, previous depth level is required. So, this part cannot be parallelized because of data dependence. But at each depth level there is plenty of roam for parallelism and can be parallelized. In case of CUDA programming, each thread calculate a single pixel on the next depth plan. After calculating the next depth plan, we can calculate the depth row by summing over all the frequencies and calculating all the depth rows results in the final migrated image. The poststack depth migration is implemented in CUDA and its performance is evaluated with the sequential code with different problem sizes.       
\end{abstract}

\IEEEpeerreviewmaketitle

\section{Introduction}
Natural resources like oil, gas, coal and other minerals are important for life. These resources are buried inside earth (both land and marine). To explore these resources we need a clear image of the earth sub-surface \cite{Yalmez}. This image can be obtained using method of reflection seismology. In this method, artificial seismic waves are generated and these waves are reflected from different geological layers which are recorded by receivers (e.g. geophones and hydrophones). The recorded data is in raw form and it needs further processing to obtain the image of earth sub-surface \cite{Yalmez,bording1997seismic}. 

Digital signal processing has played important role in many applications. Some examples are sonar, radar, medical, communications and seismology \cite{dudgeon1990multidimensional}. The actual application of signal processing in seismology began with the work of Geophysical Analysis Group at the Massachusetts Institute of Technology (MIT) in the era 1960-65, it was historical milestones in seismic data processing \cite{buttkus2000spectral}.

To obtain the image of earth subsurface some preprocessing is required to attenuate noise accompanying the data. By using an imaging technique, the time traces of the preprocessed shot records are transformed into depth traces. For this process a earth model is required and the imaging technique is called \textit{migration}.  After the imaging, a geologist can interpret the migrated section and can identify the layers and structures in the earth subsurface model. For instance, this interpretation can be used to make a decision about the position of a future borehole. If an error is made in the imaging technique and interpretation based on it is wrong, the borehole will miss its target. Therefore, a good quality of imaging technique (or seismic migration) is important. 

\textit{Seimic Migration} can be performed in different domains and in each domain there are number of algorithms. There exists many migration (extrapolation) algorithms. One migration method is frequency-wavenumber ( $f-k_x$) algorithm \cite{Yalmez,Hale-mig1}. In this method the data in time-space ($t-x$) domain is first transformed into frequency-wavenumber ($f-k_x$) domain. 
At each frequency sample, a complex-valued FIR filters are applied in the wavenumber-response domain. Another method is frequency-space ($f-x$), in which the data is transformed into frequency-space domain before migration, only the time axis is transformed to the frequency. In this method FIR extrapolation filters (known as extrapolators) are applied in space domain via convolution \cite{Hale-mig1,Holberg,Thorbecke-thesis}. By using convolution, each designed filter output can be calculated independently (or in parallel). In addition, this method can easily be extended to 3-D depth migration. This method can accurately migrate the one-way wavefields through strong laterally varying media. 

In \cite{liu20133d}, 3D reverse time migration is implemented using GPGPU. Reverse time migration is well-know seismic migration techniche but computationally expensive. Seismic migration for both time and depth is also implemented. There is huge performance gain for implementing Kirchhoff prestack time and depth migration on a cluster of 64 GPUs and 256 CPU cores \cite{panetta2012accelerating}. Nowadays, clusters are available with the GPU; so, its better to utilize all the resources. Some other previous work done for seismic signal processing (in different domains) using GPU (CUDA programming) can be found in \cite{knibbe2015reduction,wang2010accelerating,moussa2009seismic}. 

In the next section, a brief literature review is presented. After that, the implementation of poststack depth migration is discussed and followed by the performance evaluation with the results from sequential code and CUDA code. Finally, the paper is concluded. 
\section{Literature Review}
The earliest migration technique were graphical and was based on geometrical ideas developed. In 1954 Hagedoorn describes the process of seismic migration in terms of propagating wavefronts and tries to avoid the use of non-physical ray paths \cite{hagedoorn1954process}. Later on Huygens-Fresnel argued that the beam between source and receiver is at least a half wavelength wide, therefore, rather than rays it is better to work with propagating wavefronts. Huygens' principle is basis of migration \cite{Yalmez,cloerbout1982imaging}.

In seismic migration, wave propagation effects, can correctly determine the reflection points of the subsurface structures \cite{robinson1980geophysical}. Migration can be defined as the process of reconstructing a seismic section so that the reflection events are repositioned under their correct surface location at their correct vertical reflection (time or depth) location \cite{Yalmez,kearey2013introduction}. The migration process removes the distorting effects of dipping reflectors from the seismic sections. In addition it also removes the diffracted arrivals which are resulted from sharp lateral discontinuities \cite{dudgeon1984multidimensional,karam1997efficient}.

Seismic imaging in the $f-x$ domain is among the attractive methods for imaging the earth subsurface structures \cite{Yalmez,Hale-mig1,Holberg}. This method is implemented via spatial convolution to obtain seismic images \cite{Yalmez,bhardwaj1999parallel}, so each output sample can be computed independently, whenever parallel implementation is possible. Most importantly, one-way wave extrapolation in the $f-x$ domain can accurately image the subsurface with strong lateral varying medium. For strong varying medium, short length filters are required. Also, the short length filters will reduce the computational cost of convolution. At the same time, to accommodate high propagation angles of wavefields, large length filters become desirable. Hence, the design problem can be treated as an optimization problem between the two trade-offs \cite{Thorbecke-geo,Wail-Geophysics}. 

The explicit $f-x$ migration can be easily extended to the 3-D depth migration, which requires 2-D filters as extrapolators. The $f-x$ extrapolation, can accurately image the laterally varying materials by using $N$ length FIR digital filters. In the $f-x$ extrapolation, the seismic wavefields are spatially sampled, i.e. $u(x_i,e^{jw_r},z_k)$ from depth $z_k$ to $z_{k+1}=z_k+\triangle{z}$ and the process can be performed independently for each frequency $w_r$ by spatial convolution with a pre-designed filter \cite{hale1991stable}.

\begin{equation}\label{eq:extrapolate}
u({x_i},{e^{j{w_r}}},{z_{k + 1}}) = \sum\limits_{n = ( - N + 1)/2}^{(N - 1)/2} {h[n]u({x_{i - n}},{e^{j{w_r}}},{z_k})}
\end{equation}

\vspace*{0.65 cm}

In Equation \ref{eq:extrapolate}, it is shown how to apply the designed $h[n]$ filter response to extrapolate for each depth level. This process is recursive in nature as shown by the equation \ref{eq:extrapolate}. Due to this recursive nature of algorithm the extrapolation process can be unstable, so accurate designs are required to avoid instability. In the equation \ref{eq:extrapolate}, $x_i=i\Delta{x}$ and $z_k=k\Delta{z}$ for all $i$ and $k \in {\mathbb Z}$ (set of integers).. This method can be easily extended to 3-D Seismic Migration. When explicit depth migration is treated as filtering process it resembles convolution and we can compute each sample independently. This method is less expensive and can handle lateral variations in velocity.

\begin{algorithm}[b]
	\caption{Explicit $f-x$ poststack depth migration}
	\label{poststack}
	\begin{algorithmic}[1]
		\Procedure{poststack} {}

		\textbf{Input} shot gather $i$ in $(t-x)$
		
		Fourier transform of shot gather along ($t-axis$)
		
		\For{each depth level}
		
		\For{each freqeuency}
		
		\For{each space}
		
		Look-up Table (Predesigned 1-D FIR Filters)
		
		Downgoing wave extrapolation
		
		\EndFor
		
		\EndFor 
	
		Calculate depth row by summing over frequencies
		\EndFor
			
		Get the final depth migrated $(z-x)$ section
		\EndProcedure
	\end{algorithmic}
\end{algorithm}

\begin{figure*}[t]
	\centering
	\centerline{\includegraphics[width=12cm]{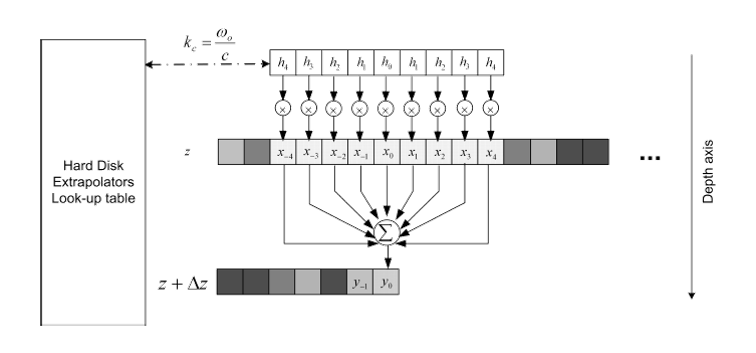}}
	\caption{Explicit depth wavefield extrapolation using 1-D FIR filters (courtesy of \cite{mousa2014imaging}).}
	\label{fig_filter}
\end{figure*}

\section{Implementation}
In this section we discuss the implementation of Poststack depth migration in $f-x$ domain. In the Algortihm \ref{poststack}, the shot gather is first transformed from $t-x$ to $f-x$ by taking the fourier transform along the time-axis. In order to calculate the image of the subsurface, we need to extrapolate the raw data based on the wave-equation. For each depth level, we repeat the process the process of extrapolation in $f-x$ domain. It simply a filtering process. This filtering process corresponds to the solution of wave-equation in frequency-wavenumber ($f-k_x$) domain. From previous depth level, next depth level is calculated. The depth level ($z-x$) is obtained by summing over the all frequencies. The cutoff of the filter depends on wavenumber (which itself depends on frequency and velocity) which can vary from point to point. So, the filter cutoffs may vary for each point. Designing filters itself is optimization problem, it takes sometime. In order to speed-up the algorithm, we design the filters in advance and store it on the disk. Based on the filters' cutoff, the required filter bank is accessed. In Figure \ref{fig_filter}, accessing the filter and applying it to the previous plan to get a pixel on the next plan is shown. We repeat the process for all depth levels/rows and for each frequency to get the final migrated image of the subsurface.    

In order to implement it on CUDA, two kernels are defined. First kernel is to calculate the next plan $f-x$ from the previous using the extrapolation filters. Second kernel is to calculate the depth row from this new plan by summing over the frequency axis. Each thread is responsible for calculating the next plan pixel. We mapped the space-axis to the blocks and frequency-axis to the threads. When the next plan is calculated, we need synchronization of threads for all the pixel. So, when the next plan is calculated by a kernel for next plan, the kernel exits in order to ensure synchronization. Now the second kernel accumulates along the frequency axis and assign it to the depth row. Each depth row represents a row in the final migrated $z-x$ section. This $z-x$ section is the final image of the earth subsurface.  

\begin{table}[b!]
	\renewcommand{\arraystretch}{1.2}
	\centering
	\caption{Execution time (minutes) of different matrix sizes.}
\begin{tabular}{c c c c c}
\hline
Problem Size & & CPU & & GPU\\
\hline
 \hline
    $128^2$ & & 0.07 & & 0.66 \\
 $256^2$ & & 0.32 & & 0.94 \\
 $512^2$ & & 1.34 & & 0.95 \\
  $1024^2$ & & 12.25 & & 1.14 \\
  $2048^2$ & & 55.87 & & 1.23 \\
  $4096^2$ & & 215.03 & & 1.47\\
\hline
\end{tabular}
	\label{tab_normal}
\vspace*{0.75 cm}
	\renewcommand{\arraystretch}{1.2}
	\centering
	\caption{Execution time (minutes) of different matrix sizes and number of threads.}
\begin{tabular}{c c c c c}
\hline
\multirow{2}{*}{} & \multicolumn{4}{c}{Number of Threads}\\
\cline{2-5}
  Problem Size & 32 & 64 & 128 & 256 \\
\hline
\hline
 $256^2$ & 0.83 & 0.71 & 0.78 & 0.99 \\
 $512^2$ & 2.36 & 1.56 & 1.22 & 1.32 \\
  $1024^2$ & 3.11 & 1.99 & 1.55 & 1.52 \\
  $2048^2$ & 4.79 & 3.10 & 2.14 & 1.93 \\
\hline
\end{tabular}
	\label{tab_anomalous}
\end{table}

\section{Performance Evaluation and Results}

\begin{figure}[t]
	\centering           
	\includegraphics[trim = 15mm 4mm 11mm 7mm, clip, width=0.5\textwidth]{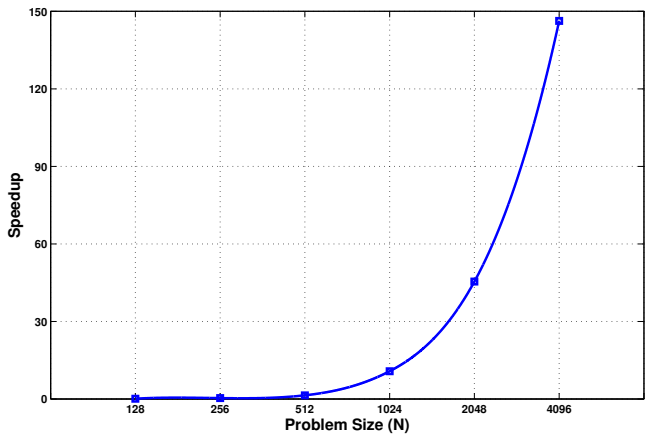} 
	\caption{Speed up over sequential program vs Problem size.} 
	\label{fig_speedup}        
\end{figure}

\begin{figure}[t]
	\centering           
	\includegraphics[trim = 15mm 4mm 11mm 7mm, clip, width=0.5\textwidth]{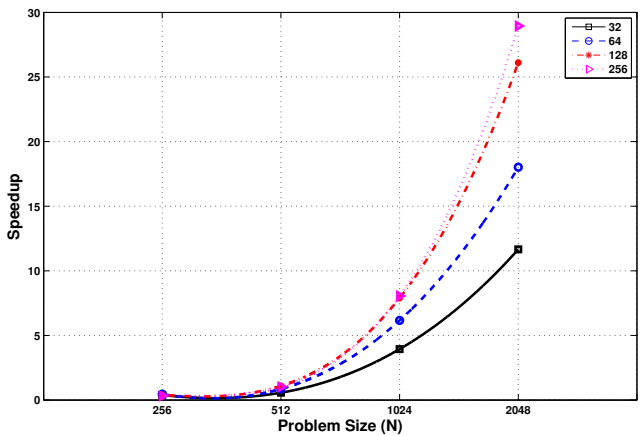} 
	\caption{Speed up with different number of threads vs Problem size.} 
	\label{fig_threads}        
\end{figure}

In this section, the performance of CUDA code is evaluated and the results from the sequential and CUDA codes are discussed. Table~\ref{tab_normal} shows a comparison of the run-time between the CPU and the GPU. We obtained a speedup of 146X for the CUDA version compared to the CPU code for the largest grid. In Figure \ref{fig_speedup}, the speedup is calculated with the problem size. As the problem size increases the CUDA code performs better. One main reason of lower speedup with smaller problem size is the overhead. So, for smaller problem size the speedup is not that high but as we increase the problem size ($N$), the parallel programs perform better and hence utilizes the GPU in better fashion. A comparison of the run-time for different number of threads and problem sizes are shown in Table~\ref{tab_anomalous}. In Figure \ref{fig_threads}, we varied the number of threads per block and plotted the speedup against the problem size ($N$). Again, the speedup increases with the problem size and we see that the highest speedup is for 256 threads per block. 

\begin{figure}[b!]
	\centering
	\centerline{\includegraphics[width=0.5\textwidth]{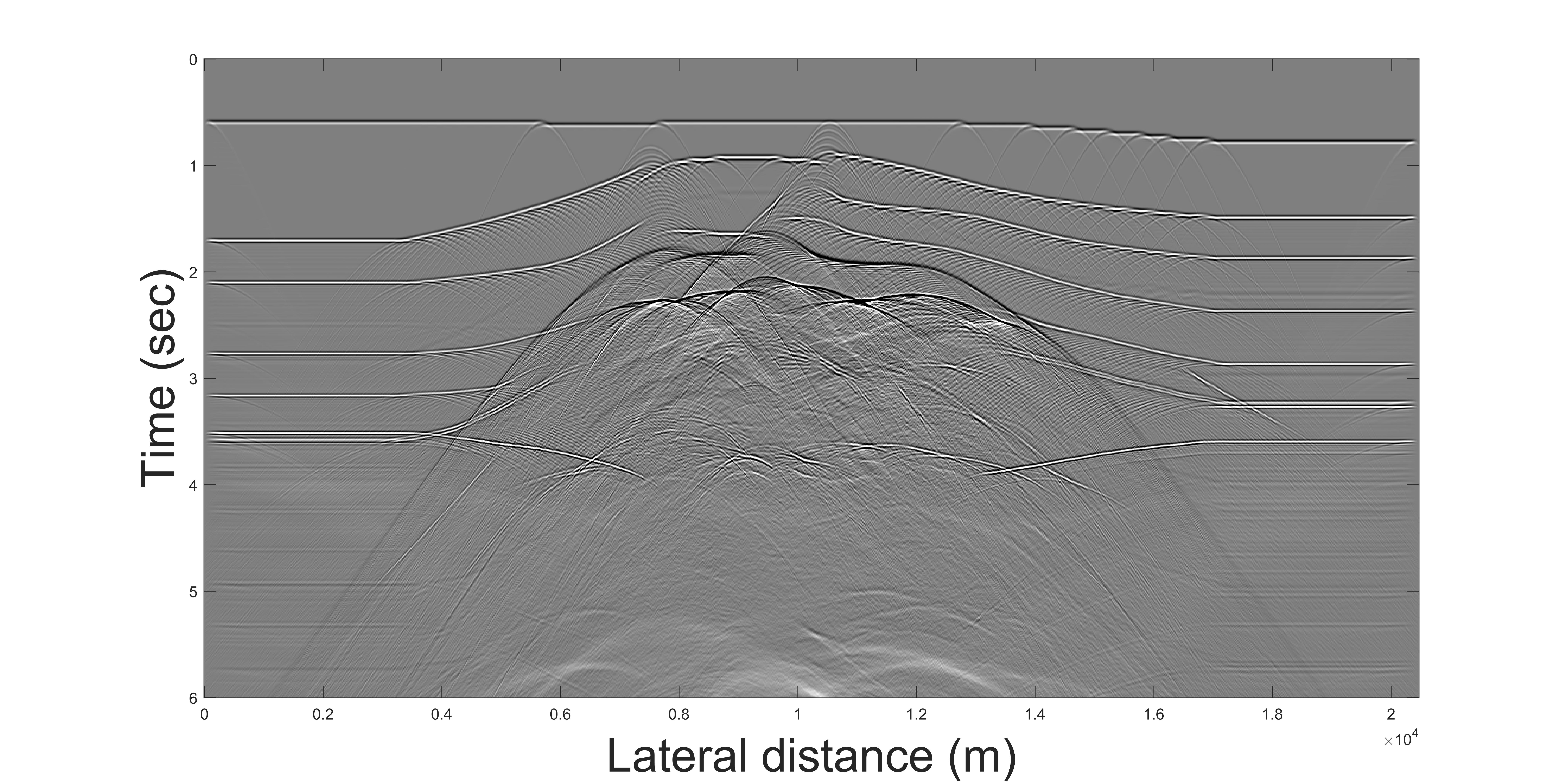}}
	\caption{SEG/EAGE salt zero-offset section.}
	\label{fig_data}
\end{figure}

\begin{figure}[t]
	\centering
	\centerline{\includegraphics[width=0.5\textwidth]{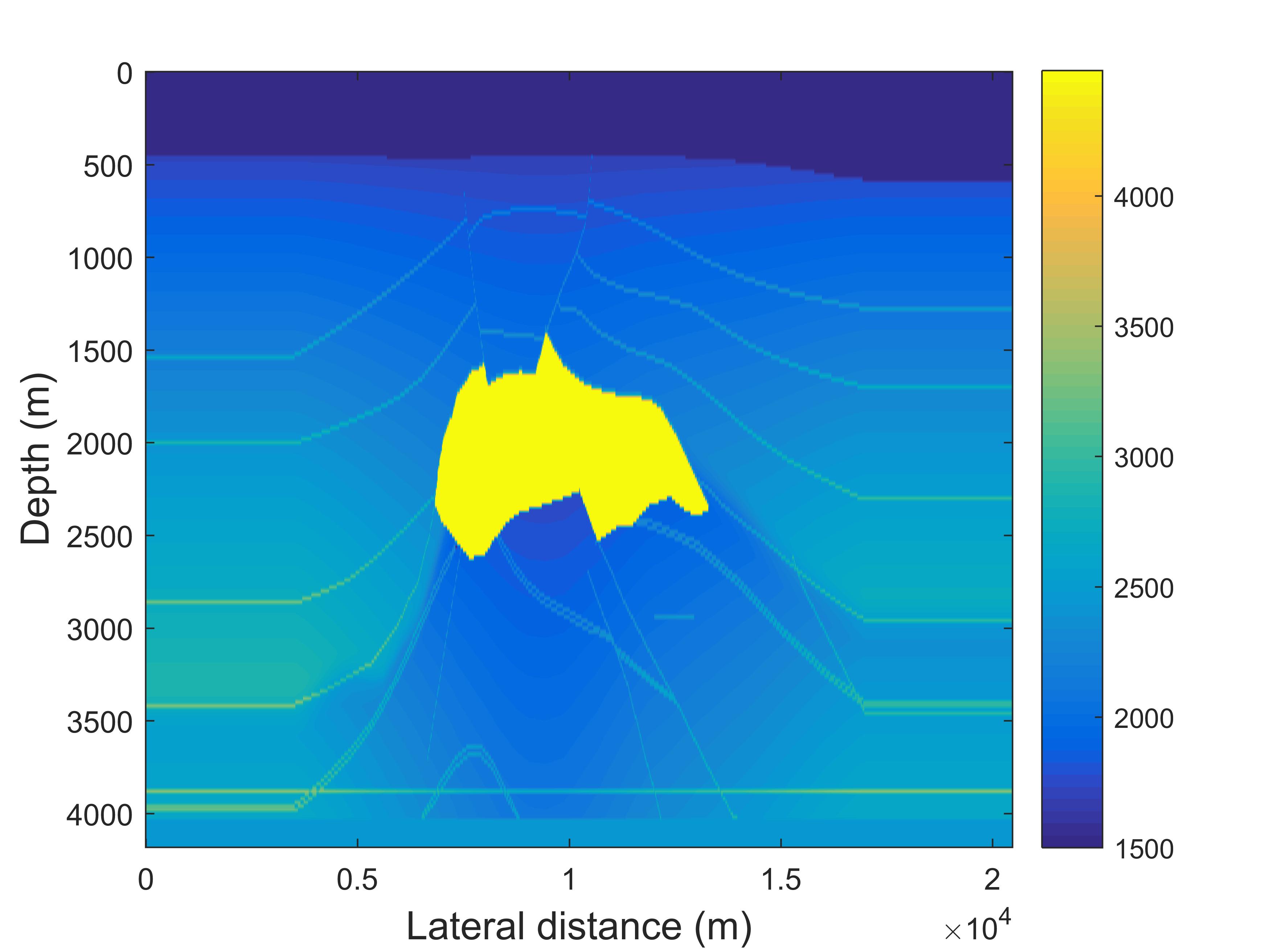}}
	\caption{SEG/EAGE salt velocity model.}
	\label{fig_model}
\end{figure}

\begin{figure}[b]
	\centering
	\centerline{\includegraphics[width=0.5\textwidth]{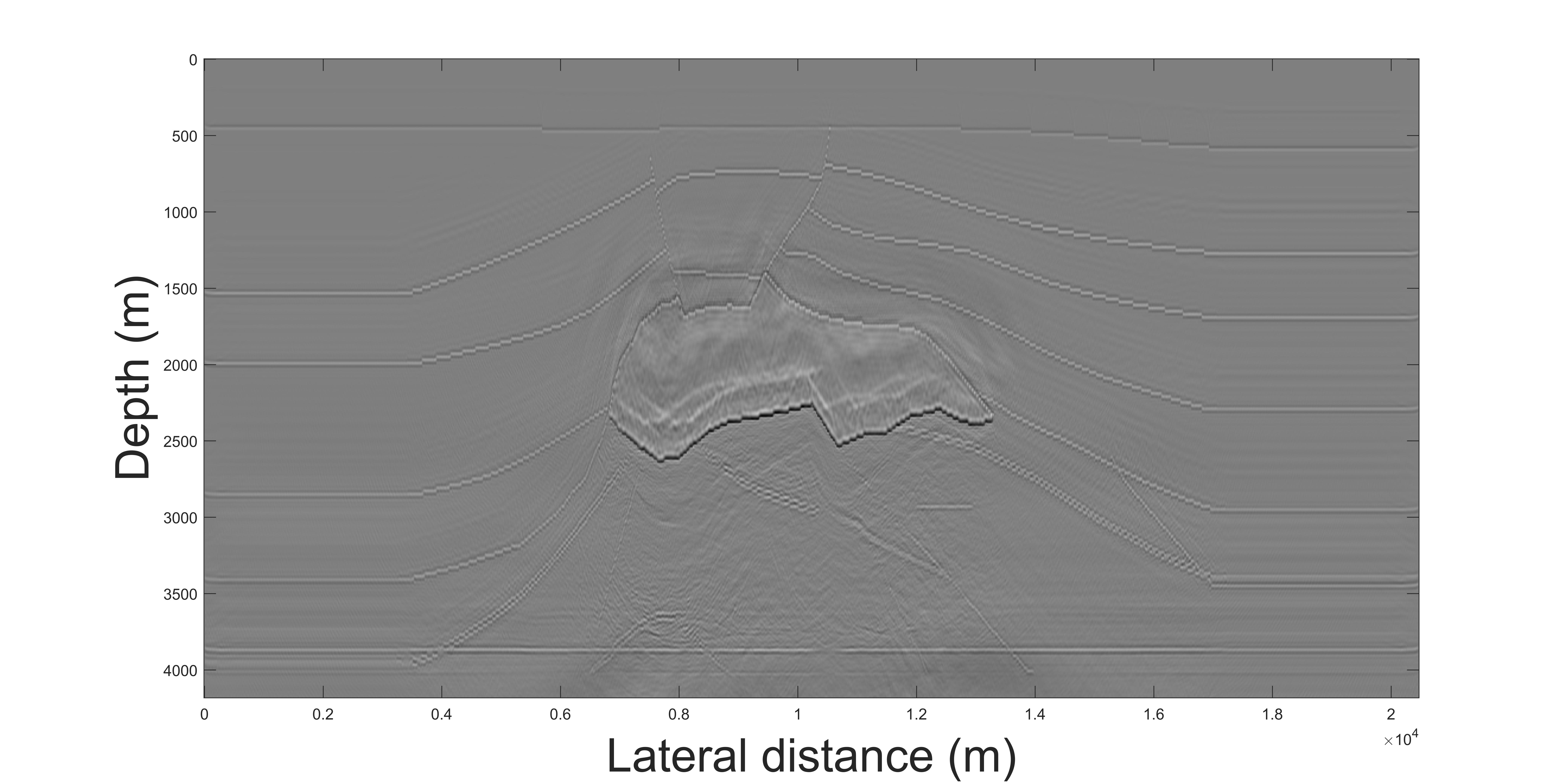}}
	\caption{Migrated section using sequential code.}
	\label{fig_result}
\end{figure}

\begin{figure}[b!]
	\centering
	\centerline{\includegraphics[width=0.5\textwidth]{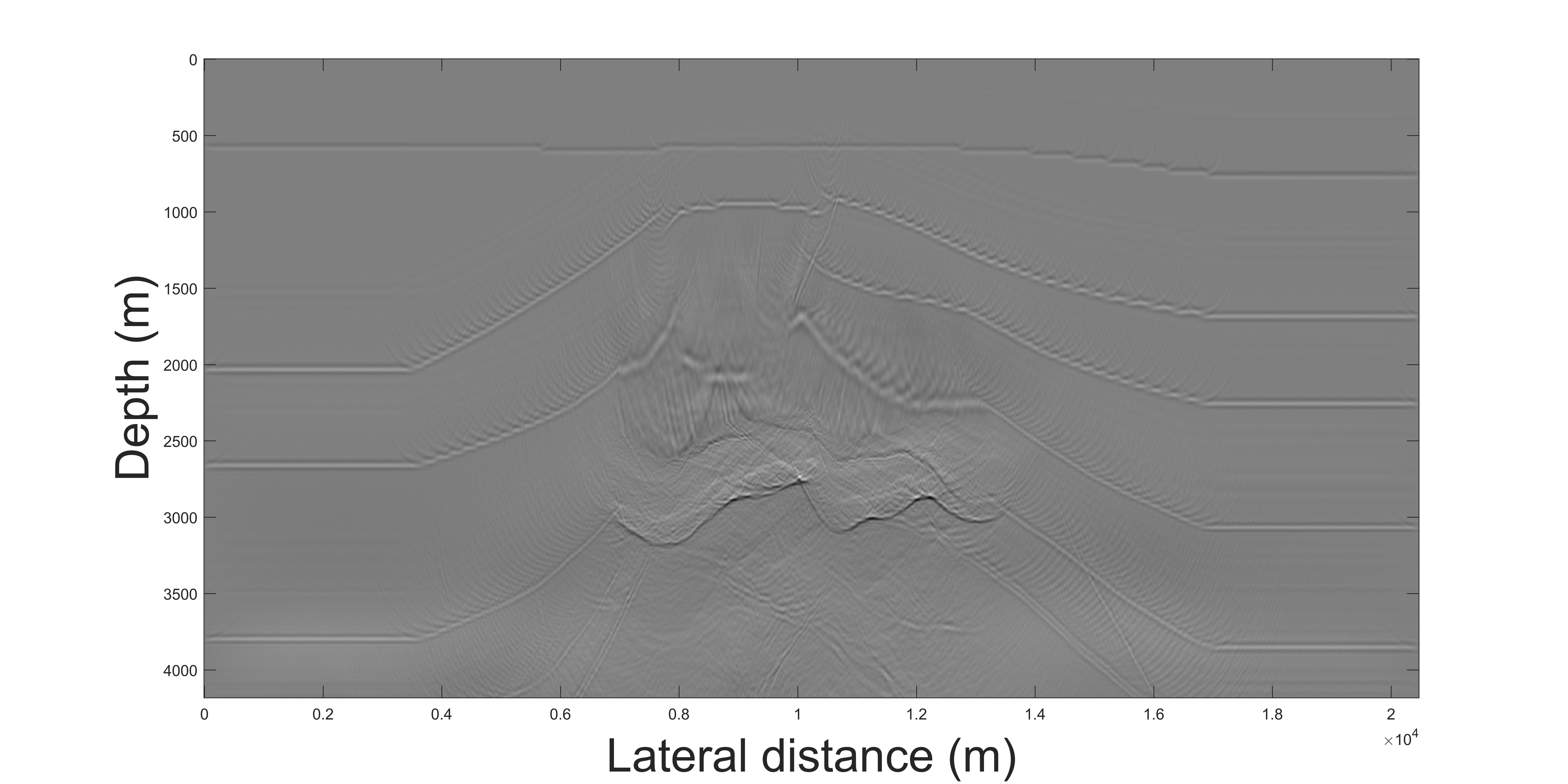}}
	\caption{Migrated section using CUDA code.}
	\label{fig_output}
\end{figure}

In this paper, we used benchmark SEG/EAGE salt model and its corresponding zero-offset raw data. In Figure \ref{fig_data}, the zero-offset data is shown. The zero-offset data is obtained from the field data after some processing. In Figure \ref{fig_model}, SEG/EAGE salt model is shown. In seismic signal processing, salt models are considered complex because of high velocity contrast and imaging beneath the salt is extremely difficult task. In Figures \ref{fig_result} and \ref{fig_output} both the migrated images from sequential code and CUDA code are compared. It seems that there is slight error in the final migrated images. We didn't achieve $100\%$ correctness. But the depth migrated section is enough close to the original image.  

\section{Conclusion}
In conclusion, the performance of the GPU is increasing with the problem size for this particular problem. As there is parallel in the algorithm at the same plan level; so, it is helpful to parallel the algorithm. At depth levels, the migration process is totally dependent on the previous depth plan. Thus, it is important to wait for all the thread to finish and synchronize the parallel program at that level. In the future, this work can be easily extended to the prestack depth migration, which is relatively more computationally expensive. Another extension is the block level synchronization. We used exit from the kernel for synchronization, it can also be achieved with some other alternative.   

\section{Acknowledgment}

The authors would like to thank King Fahd University of Petroleum and Minerals (KFUPM) for supporting this research and providing the computing facilities.

\bibliographystyle{IEEEtran}
\bibliography{main}

\end{document}